\documentclass[twocolumn,aps,prb,showpacs,superscriptaddress,amsmath,amssymb]{revtex4}
\usepackage[latin1]{inputenc}
\usepackage{graphicx}% Include figure files
\usepackage{dcolumn}% Align table columns on decimal point
\usepackage{bm}% bold math
\usepackage{color}

\begin{document}

%\bibliographystyle{prsty}%{apsrev}

%%%%%%%%%%%%%%%%%%%%%%%%%%%%%%%%%%%%%%%%%%%%%%%%%%%%%%%%%%%%
%%%%%%%%%%%%%%%%%%%%%%%%%%%%%%%%%%%%%%%%%%%%%%%%%%%%%%%%%%%%
%%%%%%%%%%%%%%%%%%%%%%%%%%%%%%%%%%%%%%%%%%%%%%%%%%%%%%%%%%%%
%%%%%%%%%%%%%%%%%%%%%%%%%%%%%%%%%%%%%%%%%%%%%%%%%%%%%%%%%%%%
%%%%%%%%%%%%%%%%%%%%%%%%%%%%%%%%%%%%%%%%%%%%%%%%%%%%%%%%%%%%

\title{Molecular prototypes for spin-based CNOT quantum gates}

\author{F. Luis}
\email{fluis@unizar.es} \affiliation{Instituto de Ciencia de
Materiales de Arag\'on, C.S.I.C. - Universidad de Zaragoza, and Dpto.
de F\'{\i}sica de la Materia Condensada, Universidad de Zaragoza, E-50009 Zaragoza, Spain}%Lines break automatically or can be forced %with \\
\author{A. Repoll\'es}%
\affiliation{Instituto de Ciencia de Materiales de Arag\'on,
C.S.I.C. - Universidad de Zaragoza, and Dpto.
de F\'{\i}sica de la Materia Condensada, Universidad de Zaragoza, E-50009 Zaragoza, Spain}%
\author{M. J. Mart\'{\i}nez-P\'erez}%
\affiliation{Instituto de Ciencia de Materiales de Arag\'on,
C.S.I.C. - Universidad de Zaragoza, and Dpto.
de F\'{\i}sica de la Materia Condensada, Universidad de Zaragoza, E-50009 Zaragoza, Spain}%
\author{D. Aguil\`{a}}%
\affiliation{Departament de Química Inorgànica, Universitat de Barcelona, Diagonal 647, 08028, Barcelona, Spain}%
\author{O. Roubeau}%
\affiliation{Instituto de Ciencia de Materiales de Arag\'on,
C.S.I.C. - Universidad de Zaragoza, and Dpto.
de F\'{\i}sica de la Materia Condensada, Universidad de Zaragoza, E-50009 Zaragoza, Spain}%
\author{D. Zueco}%
\affiliation{Instituto de Ciencia de Materiales de Arag\'on,
C.S.I.C. - Universidad de Zaragoza, and Dpto.
de F\'{\i}sica de la Materia Condensada, Universidad de Zaragoza, E-50009 Zaragoza, Spain}%
\author{M. Evangelisti}%
\affiliation{Instituto de Ciencia de Materiales de Arag\'on,
C.S.I.C. - Universidad de Zaragoza, and Dpto.
de F\'{\i}sica de la Materia Condensada, Universidad de Zaragoza, E-50009 Zaragoza, Spain}%
\author{A. Cam\'on}%
\affiliation{Instituto de Ciencia de Materiales de Arag\'on,
C.S.I.C. - Universidad de Zaragoza, and Dpto.
de F\'{\i}sica de la Materia Condensada, Universidad de Zaragoza, E-50009 Zaragoza, Spain}%
\author{J. Ses\'e}%
\affiliation{Instituto de Nanociencia de Arag\'on, Universidad de Zaragoza, and Dpto. de
F\'{\i}sica de la Materia Condensada, Universidad de Zaragoza, E-50018 Zaragoza, Spain}%
\author{L. A. Barrios}%
\affiliation{Departament de Química Inorgànica, Universitat de Barcelona, Diagonal 647, 08028, Barcelona, Spain}%
\author{G. Arom\'{\i}}%
\email{guillem.aromi@antares.qi.ub.es}\affiliation{Departament de Química Inorgànica, Universitat de Barcelona, Diagonal 647, 08028, Barcelona, Spain}%

%%%%%%%%%%%%%%%%%%%%%%%%%%%%%%%%%%%%%%%%%%%%%%%%%%%%%%%%

\date{\today}% It is always \today, today,
             %  but any date may be explicitly specified

%%%%%%%%%%%%%%%%%%%%%%%%%%%%%%%%%%%%%%%%%%%%%%%%%%%%%%%%%%%%
%%%%%%%%%%%%%%%%%%%%%   ABSTRACT         %%%%%%%%%%%%%%%%%%%
%%%%%%%%%%%%%%%%%%%%%   ABSTRACT         %%%%%%%%%%%%%%%%%%%
%%%%%%%%%%%%%%%%%%%%%   ABSTRACT         %%%%%%%%%%%%%%%%%%%
%%%%%%%%%%%%%%%%%%%%%   ABSTRACT         %%%%%%%%%%%%%%%%%%%
%%%%%%%%%%%%%%%%%%%%%   ABSTRACT         %%%%%%%%%%%%%%%%%%%
%%%%%%%%%%%%%%%%%%%%%%%%%%%%%%%%%%%%%%%%%%%%%%%%%%%%%%%%%%%%

\begin{abstract}
We show that a chemically engineered structural asymmetry in [Tb$_{2}$] molecular clusters renders the two weakly coupled Tb$^{3+}$ spin qubits magnetically inequivalent. The magnetic energy level spectrum of these molecules meets then all conditions needed to realize a universal CNOT quantum gate.
\end{abstract}


%\keywords{Suggested keywords}%Use showkeys class option if keyword
                              %display desired
\maketitle

%\tableofcontents

%%%%%%%%%%%%%%%%%%%%%%%%%%%%%%%%%%%%%%%%%%%%%%%%%%%%%
%%%%%%%%%%%%%%%%%%%%   TEXT    %%%%%%%%%%%%%%%%%%%%%%
%%%%%%%%%%%%%%%%%%%%   TEXT    %%%%%%%%%%%%%%%%%%%%%%
%%%%%%%%%%%%%%%%%%%%   TEXT    %%%%%%%%%%%%%%%%%%%%%%
%%%%%%%%%%%%%%%%%%%%   TEXT    %%%%%%%%%%%%%%%%%%%%%%
%%%%%%%%%%%%%%%%%%%%   TEXT    %%%%%%%%%%%%%%%%%%%%%%
%%%%%%%%%%%%%%%%%%%%%%%%%%%%%%%%%%%%%%%%%%%%%%%%%%%%%

\begin{figure}[ht!]
%\hspace*{-2.em}
\resizebox{7 cm}{!}{\includegraphics{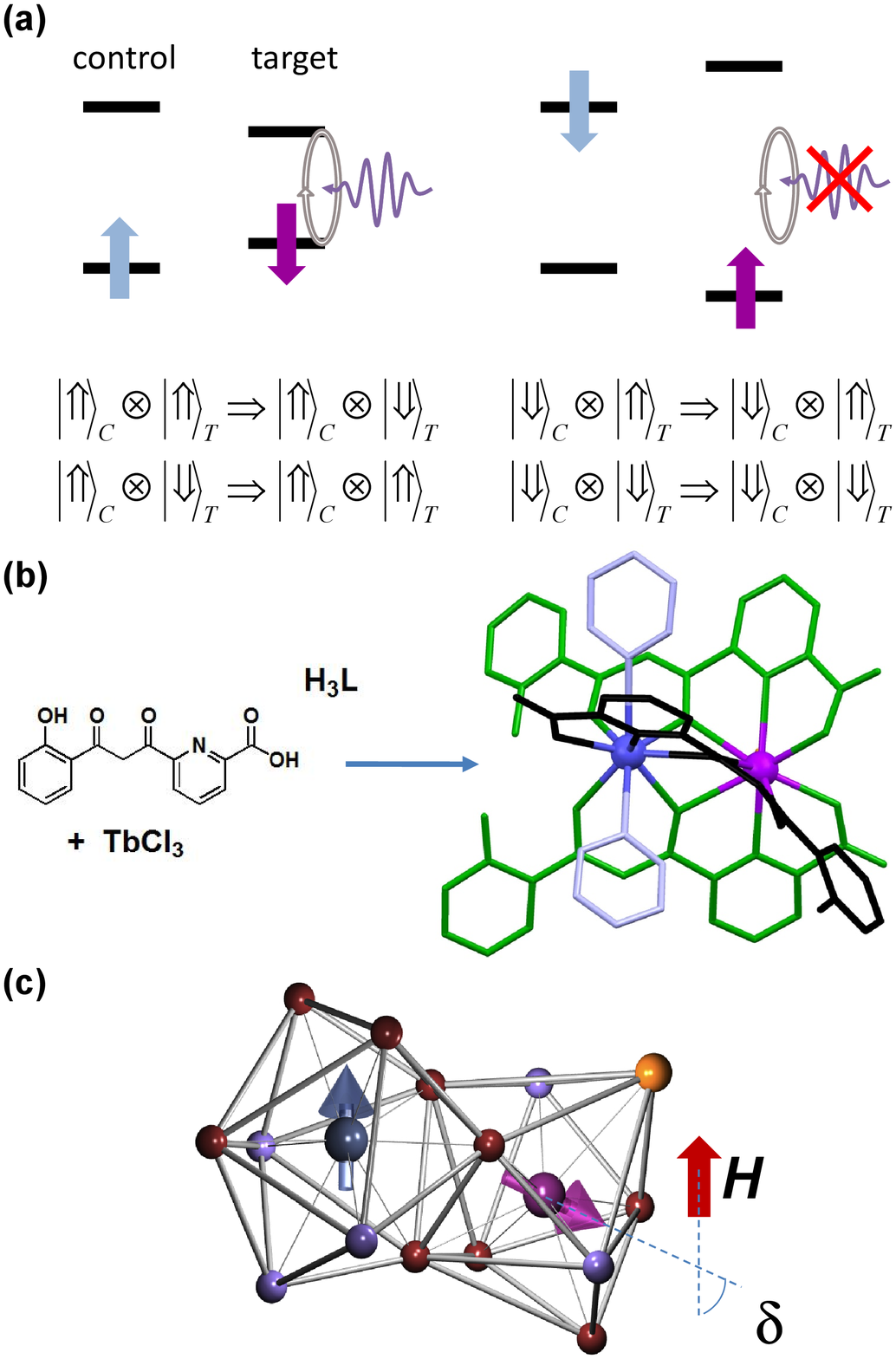}}
%\vspace*{-4.ex}
\caption{(Color online) (a) Schematic illustration of a quantum CNOT operation on two coupled spin qubits. (b) Targeted synthesis\cite{Aguila10} of an asymmetric [Tb$_{2}$] complex using three H$_{3}$L ligands (depicted in green or black colors in the complex). (c) The resulting N$_{2}$O$_{6}$Cl and N$_{3}$O$_{6}$ coordination polyhedra around the two Tb$^{3+}$ ions strongly differ, exhibiting C$_{1}$ symmetry and C$_{4v}$ symmetry, respectively (see \onlinecite{EPAPS} for details), and induce a misalignment between the two local anisotropy axes. Color code: "control" Tb, purple; "target" Tb, dark blue; Cl, orange; O, red; N, light blue. }
%% \vspace*{-3.ex}
\label{Molecule}
\end{figure}

Quantum computation\cite{Nielsen00,Ladd10} relies on the physical realization of quantum bits and quantum gates. The former can be in any of two distinguishable and addressable states, denoted here as spin up $|\Uparrow \rangle$ and spin down $|\Downarrow \rangle$, but also, as opposed to their classical counterparts, in any arbitrary linear superposition of these. The latter involve controlled operations on two coupled qubits.\cite{Nielsen00} The universal controlled-NOT (CNOT) gate is the archetype of such a controlled operation. It flips the target qubit depending on the state of the other, or control qubit (see Fig. \ref{Molecule}(a)). In terms of its material realization, this definition implies that each of the two qubits should respond inequivalently to some external stimulus, {\it e.g.} electric or magnetic fields.

Solid-state candidates to build these basic ingredients include superconducting circuits,\cite{Clarke08,Plantenberg07,DiCarlo09} single spins in semiconductors,\cite{Burkard99,Jelezko04,Hanson08} and, more recently, molecular nanomagnets.\cite{Leuenberger01,Tejada01,Troiani05,Lehmann07,Stamp09,Ardavan09} The latter are attractive for scalability, since macroscopic arrays of identical molecular qubits can be prepared via relatively simple chemical methods.\cite{Miyake09,Mannini10} State-of-the-art achievements include the measurement and minimization of decoherence rates\cite{Ardavan07,Bertaina08,Schlegel08} and the synthesis of mutually interacting qubit pairs.\cite{Timco09,Candini09} However, the realization of a universal CNOT quantum gate inside a molecular cluster remains an outstanding challenge.\cite{Ardavan09} Here, we show that asymmetric [Tb$_{2}$] molecular clusters show also the magnetic asymmetry that is required for the realization of spin-based CNOT gates.

Lanthanide ions are promising candidates for encoding quantum information.\cite{Bertaina07} The magnetic energy levels are split by the crystal field generated by surrounding atoms, which often results in a well-isolated ground state doublet. Each metal ion behaves then as an effective (two-state) spin-$1/2$, therefore providing a good realization of qubit states. For the realization of a spin-based qugate, it seems therefore natural to look for simple molecules, made of just two weakly-coupled lanthanide qubits. However, the synthesis of asymmetric molecular dimers is not straightforward, as Nature tends to make them symmetric. We propose a simple solution that exploits the ability of chemical design to finely tune the internal molecular structure. The chemical reaction is sketched in Fig.~\ref{Molecule}(b). By the use of an adequate asymmetric organic ligand\cite{Aguila10} H$_{3}$L, we synthesize a dinuclear complex of Tb$^{3+}$ ions, hereafter briefly referred to as [Tb$_{2}$], in which the metallic dimer is wrapped by three ligands. Therefore, each metal ion in the molecule is in a different coordination environment (see Fig. \ref{Molecule}(c)). In the following, we describe physical experiments that enable us to verify that this molecular cluster fulfills the basic conditions to act as a CNOT, namely the appropriate definition of the two qubits, the existence of a weak coupling between them, and the magnetic asymmetry.

Magnetic and calorimetric measurements were performed on powdered specimens. Above $1.8$ K, ac ($h_{\rm ac}=4$~Oe) susceptibility and magnetization data were measured with a commercial SQUID magnetometer. Ac susceptibility data were also measured with a $\mu$SQUID susceptometer,\cite{Martinez10} operating in the $13~{\rm mK}<T<1.5$~K temperature and $0.01$~Hz up to $1$~MHz frequency ranges. The amplitude of the ac magnetic field was, in this case, $h_{\rm ac} = 1$~mOe. Dc-magnetization measurements below $2$~K were performed using a homemade Hall microprobe installed in a dilution refrigerator. Heat capacity measurements down to $\approx 0.35$~K were performed using the relaxation method by means of a commercial setup.

\begin{figure}[ht!]
%\hspace*{-2.em}
\resizebox{7 cm}{!}{\includegraphics{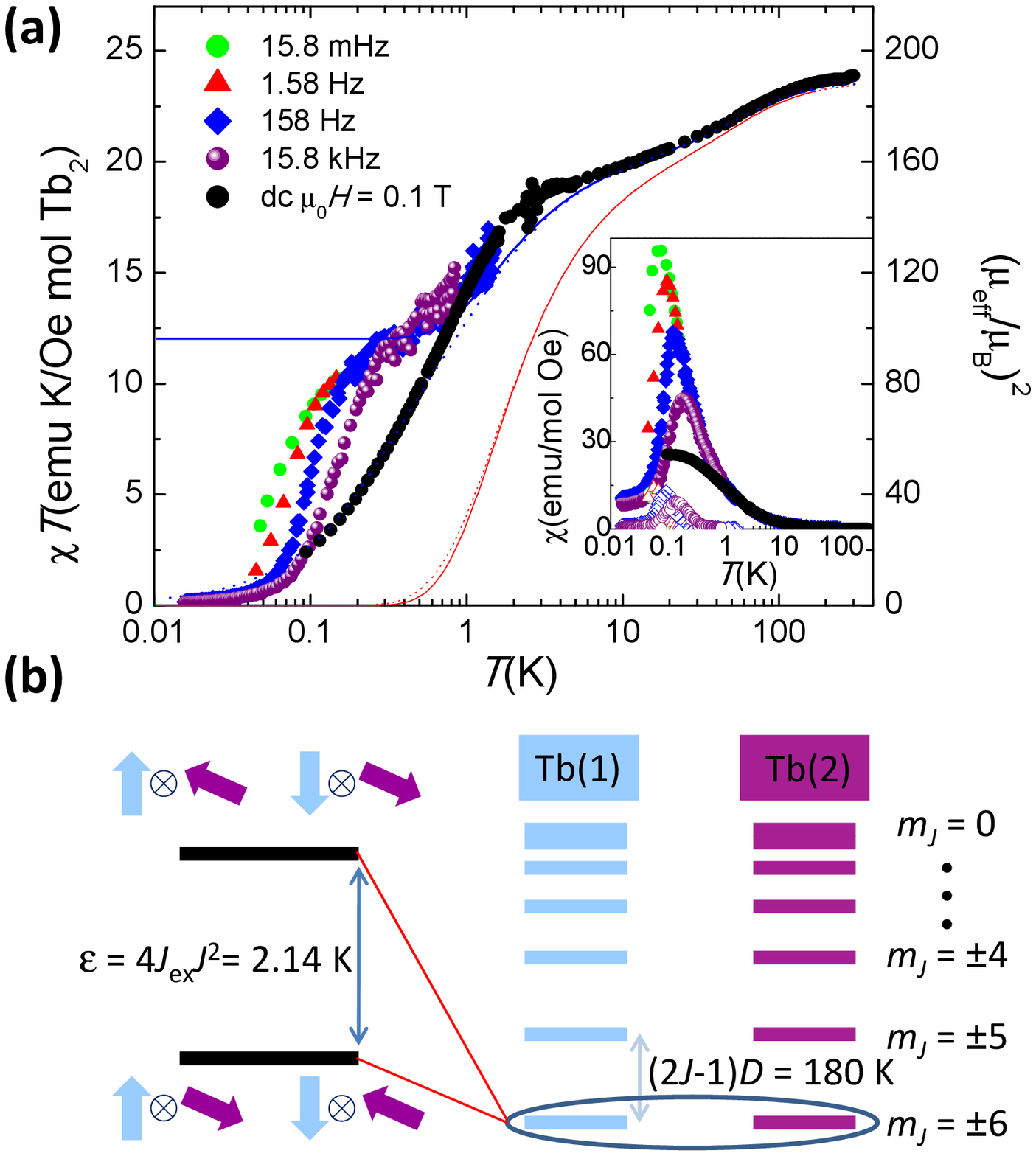}}
%\vspace*{-4.ex}
\caption{(Color online) (a) Ac susceptibility of [Tb$_{2}$] measured on a powdered sample for different frequencies of the alternating excitation field. Main panel: $\chi^{\prime}T$ product (left axis) from which $\mu_{\rm eff}$ is determined (right axis). Inset: $\chi^{\prime}$ (solid symbols) and $\chi^{\prime \prime}$ (open symbols). Dc susceptibility data measured under a constant magnetic field $\mu_{0} H = 0.1$ T are also shown. The lines represent least square fits of the ac (solid lines) and dc (dotted lines) susceptibilities based on the spin Hamiltonian (\ref{Hamiltonian1}) for collinear ($\delta=0$, red thin lines) and noncollinear ($\delta=66^{\circ}$, blue thick lines) anisotropy axes. (b) Zero-field energy level structure of [Tb$_{2}$] derived from these fits.}
%% \vspace*{-3.ex}
\label{chivsT}
\end{figure}

The ac magnetic susceptibility $\chi$ provides direct insight on the magnetic anisotropy of the Tb$^{3+}$ ions and their mutual coupling. For the lowest frequency ($0.0158$ Hz) and above $100$ mK, the in-phase component $\chi^{\prime}$ gives the equilibrium paramagnetic response (see Fig. \ref{chivsT} (a)). In this regime, the cluster's effective magnetic moment $\mu_{\rm eff}$ can be approximately determined as $\mu_{\rm eff} \simeq (3k_{\rm B} \chi^{\prime}T/N_{\rm A})^{1/2}$. At room temperature, $\mu_{\rm eff} = 13.7(1) \mu_{\rm B}$ agrees with the effective moment of two uncoupled free ions, {\it i.e.} $\mu_{\rm eff} = g_{J} \mu_{\rm B} [2J(J+1)]^{1/2} = 13.74 \mu_{\rm B}$, where $g_{J} = 3/2$ and $J = 6$ are the gyromagnetic ratio and the total angular momentum given by Hund's rules. The drop observed below approximately $100$~K can be assigned to the thermal depopulation of magnetic energy levels split by the crystal field. The value $\mu_{\rm eff} = 12.5(1) \mu_{\rm B}$ measured between $3$ and $10$~K is close to the ``Ising'' limit $\mu_{\rm eff} = g_{J} \mu_{\rm B} 2^{1/2} J = 12.72 \mu_{\rm B}$, characteristic of two uncoupled Tb$^{3+}$ ions whose angular momenta $\vec J_1$ and $\vec J_2$ can only point either up or down along their local anisotropy axes. The magnetic anisotropy can be determined by fitting $\chi^{\prime} T$ in the intermediate temperature regime, using expressions derived\cite{EPAPS,GarciaPalacios09} for the simplest uniaxial anisotropy ${\cal H} = -D(J_{1,z}^{2} + J_{2,z}^{2})$. The fit gives $D/k_{\rm B} = 17$~K. For such a strong anisotropy, excited levels are separated by more than $180$~K from the ground state doublet, associated with the maximum projections $m_{J}=\pm J$ (see Fig. \ref{chivsT}(b)). These states provide then, for each ion, a proper definition of the qubit basis $|\Uparrow\rangle$ and $|\Downarrow\rangle$, namely:

\begin{align}
|\Uparrow\rangle &\equiv |J=6, m_J=6\rangle
\\
\qquad
|\Downarrow\rangle &\equiv |J=6, m_J=-6\rangle
\end{align}

\begin{figure}[ht!]
%\hspace*{-2.em}
\resizebox{7 cm}{!}{\includegraphics{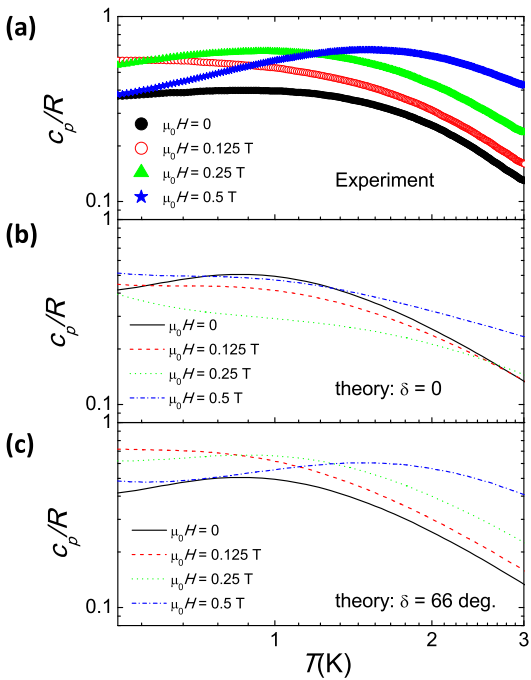}}
%\vspace*{-4.ex}
\caption{(Color online) Magnetic heat capacity of [Tb$_{2}$]. (a) Data measured on a powdered sample under different values of the applied magnetic field. (b) and (c) Theoretical predictions for collinear and noncollinear anisotropy axes, respectively.}
%% \vspace*{-3.ex}
\label{CpvsT}
\end{figure}

By further decreasing temperature below $3$~K, we observe a second drop in $\chi^{\prime}T$ (Fig.~\ref{chivsT}) that we associate with the antiferromagnetic coupling $-2J_{\rm ex} \overrightarrow{J_{1}} \overrightarrow{J_{2}}$ of the two qubits. This interpretation is corroborated by the magnetic heat capacity $c_{m,P}$, shown in Fig.~\ref{CpvsT} (see \onlinecite{EPAPS}). At zero field, $c_{m,P}$ shows a Schottky-type broad anomaly centered at $T_{\rm max}\simeq 0.9$~K, which arises from the energy splitting $\epsilon$ between antiferromagnetic ($|\Uparrow\rangle \otimes |\Downarrow\rangle$ and $|\Downarrow\rangle \otimes |\Uparrow\rangle$) and ferromagnetic ($|\Uparrow\rangle \otimes |\Uparrow\rangle$ and $|\Downarrow \rangle \otimes |\Downarrow \rangle$) states. Using the condition $k_{\rm B}T_{\rm max} = 0.42 \epsilon$, we determine the splitting $\epsilon/k_{\rm B} \simeq 4J_{\rm ex}J^{2}/k_{\rm B} \simeq 2.14$~K and, from this, the effective exchange constant $J_{\rm ex}/k_{\rm B} = -0.016(1)$~K.

Considering the value of $\epsilon$, the monotonic increase of $\chi^{\prime}$ down to $100$~mK would be puzzling unless the magnetic moments of the two Tb$^{3+}$ ions do not exactly compensate each other. As the coordination sphere determines the magnetic anisotropy, the anisotropy axes of the two ions need not be parallel to each other, but can instead make a tilting angle $\delta$, as shown in Fig. \ref{Molecule}. Because of this misalignment and the very strong anisotropy, even the states ($|\Uparrow\rangle \otimes |\Downarrow\rangle$ and $|\Downarrow\rangle \otimes |\Uparrow\rangle$) preserve a net magnetic moment. More importantly, the two ions will couple differently to an external magnetic field, i.e. their effective gyromagnetic ratios $g_{1}$ and $g_{2}$ will be different. For instance, if $H$ is applied along one of the anisotropy axes, say of qubit ``1'', $g_{1} = g_{J}$ whereas $g_{2} = g_{J} \cos{\delta}$, {\it i.e.} the field makes the two spins inequivalent, just as required for the CNOT operation.

Susceptibility and heat capacity measurements confirm that $g_{1} \neq g_{2}$. The absence of magnetic cancelation in the cluster causes the paramagnetic response observed between $10$ K and $100$~mK (see the inset of  Fig.~\ref{chivsT}). Indeed, below $10$~K $\chi^{\prime} T$ (and thus also $\mu_{\rm eff}$) is much larger than predicted for collinear anisotropy axes ($\delta=0$). In contrast, an excellent agreement is obtained for $\delta = 66^{\circ}$. We have also measured the dc-susceptibility under a magnetic field of $0.1$~T, obtaining further experimental evidence for noncollinear axes with the same values of $J_{\rm ex}$ and $\delta$ estimated above.

The same conclusion is drawn from heat capacity data measured under nonzero applied magnetic fields, shown in Fig. \ref{CpvsT}. This quantity reflects the magnetic field dependence of the energy level structure, which, in its turn, should strongly depend on $\delta$. As with the magnetic data, the results are in qualitative and quantitative agreement with the calculations made for $\delta = 66^{\circ}$.

\begin{figure}[ht!]
%\hspace*{-2.em}
\resizebox{7 cm}{!}{\includegraphics{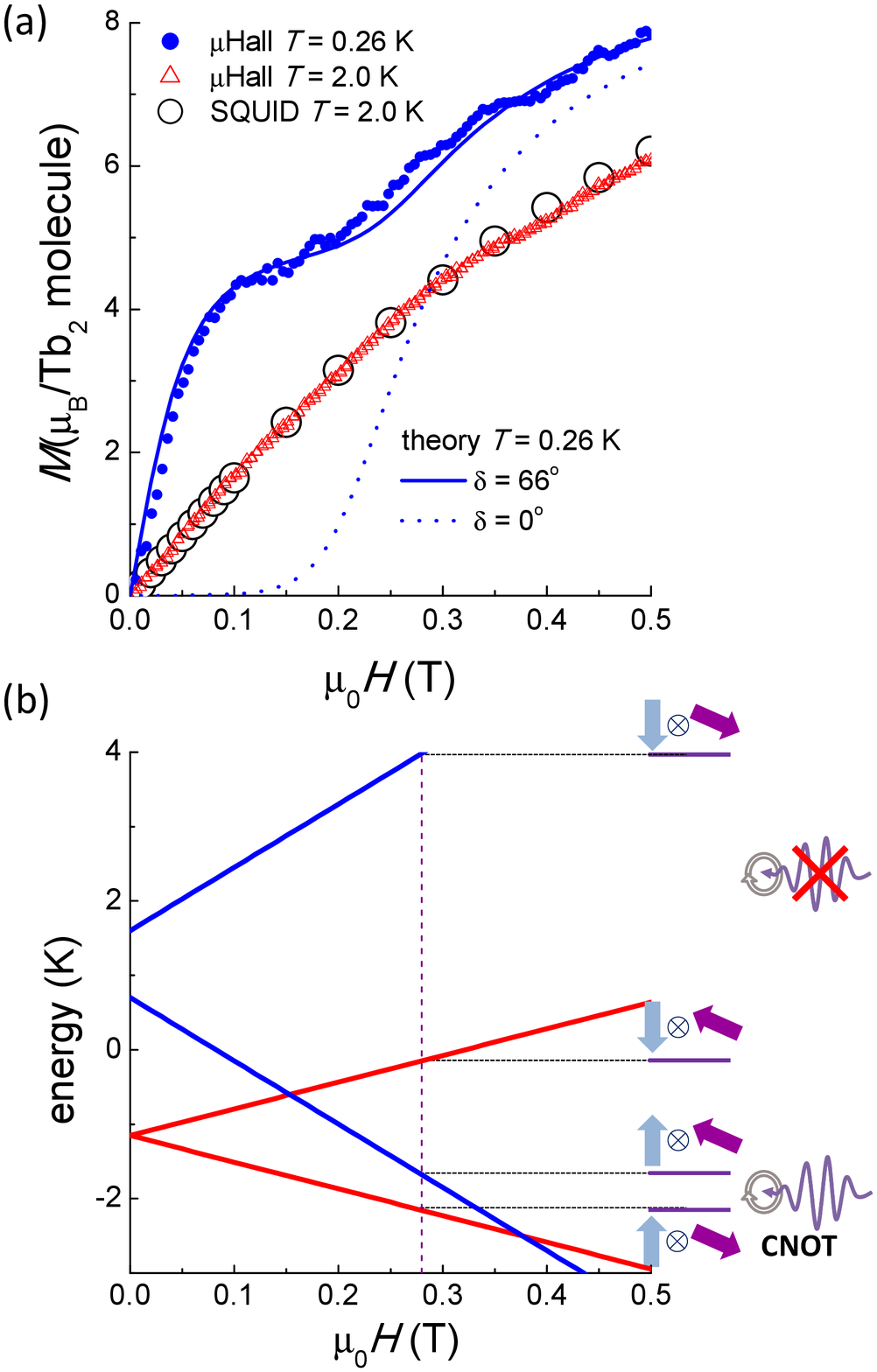}}
%\vspace*{-4.ex}
\caption{(Color online) (a) Magnetization isotherms of [Tb$_{2}$] measured at two temperatures. The data at $T=0.26$ K are compared with calculations made for collinear ($\delta=0$, dotted line) and noncollinear ($\delta=66^{\circ}$, solid line)  anisotropy axes. (b) Field-dependent magnetic energy levels of [Tb$_{2}$] calculated by numerical diagonalization of the Hamiltonian (\ref{Hamiltonian1}) for $\delta=66^{\circ}$. Only levels corresponding to magnetic nuclear states $m_{I,1} = m_{I,2} = -3/2$ are shown.}
%% \vspace*{-3.ex}
\label{MvsH}
\end{figure}

As an independent confirmation of the magnetic inequivalence of the two Tb$^{3+}$ ions, we have also measured magnetization isotherms at fixed temperatures $T=0.26$ and $2$~K. The results are shown in Fig.~\ref{MvsH}(a). The presence of a non-compensated $\mu_{\rm eff}$ has dramatic consequences at very low temperatures and for low magnetic fields, when the thermal populations of the excited ferromagnetic states ($|\Uparrow\rangle \otimes |\Uparrow\rangle$ and $|\Downarrow \rangle \otimes |\Downarrow \rangle$) become negligibly small. For $\delta=0$, the ground state of the cluster would be nonmagnetic, therefore $M \simeq 0$ until it would abruptly change at the magnetic field $\mu_{0}H \sim 0.15$~T where the energy level crossing occurs. Experimentally, we find instead a finite paramagnetic response starting already from $H=0$, which shows that the molecular ground state possess a net magnetic moment (and therefore that $\delta$ must be different from $0$ and close indeed to $66^{\circ}$). The level crossing does still give rise to an abrupt magnetization jump at a somewhat higher field, confirming the antiferromagnetic character of the exchange interactions and the value of $J_{\rm ex}$ determined from heat capacity experiments.

All experiments presented so far lead us to conclude that the essential physics is captured by a Hamiltonian containing the uniaxial anisotropy, the exchange couplings, the Zeeman energy and, finally, the hyperfine interactions with the nuclear spins $I=3/2$ of Tb. In addition, we have already anticipated that the relevant qubit states $|\Uparrow\rangle$ and $|\Downarrow\rangle$ are related to $m_J = \pm 6$. In this reduced subspace, the Hamiltonian simplifies to\cite{EPAPS}

\begin{align}
\nonumber
{\cal H} =& -2J_{\rm ex} J_{1,z} J_{2,z} - g_{1} \mu_{\rm B} H J_{1,z} - g_{2} \mu_{\rm B} H J_{2,z}\\
&+A_{J} \left(J_{1,z}I_{1,z} + J_{2,z}I_{2,z} \right)
\label{Hamiltonian1}
\end{align}

\noindent where $A_{J}/k_{\rm B} = 2.5 \times 10^{-2}$~K is the hyperfine constant.

%%%%%%%%%%%%%%%%%%%%%%%%%%%%%%%%%%%%%%%%%%%%%%%%%%%%%%%%%%
%%%%%%%%%%%%%%%%%%%%   CONCLUSIONS  %%%%%%%%%%%%%%%%%%%%%%
%%%%%%%%%%%%%%%%%%%%   CONCLUSIONS  %%%%%%%%%%%%%%%%%%%%%%
%%%%%%%%%%%%%%%%%%%%   CONCLUSIONS   %%%%%%%%%%%%%%%%%%%%%
%%%%%%%%%%%%%%%%%%%%   CONCLUSIONS   %%%%%%%%%%%%%%%%%%%%%
%%%%%%%%%%%%%%%%%%%%   CONCLUSIONS   %%%%%%%%%%%%%%%%%%%%%
%%%%%%%%%%%%%%%%%%%%%%%%%%%%%%%%%%%%%%%%%%%%%%%%%%%%%%%%%%

The Hamiltonian (\ref{Hamiltonian1}) enables us to finally discuss the performance of [Tb$_{2}$] as a CNOT qugate. The ensuing energy level spectrum is shown in Fig.~\ref{MvsH}(b) for $\delta=66^{\circ}$. For the sake of clarity, we show only levels associated with a single nuclear magnetic state of the cluster, namely with nuclear spin projections $m_{I,1} = m_{I,2} = -3/2$. The magnetic asymmetry induced by the tilting angle $\delta=66^{\circ}$, enables to univocally single out any of the desired transitions. For instance, at $\mu_{0}H=0.28$~T, only the two states $|\Uparrow\rangle_{1} \otimes |\Uparrow\rangle_{2}$ and $|\Uparrow\rangle_{1} \otimes |\Downarrow\rangle_{2}$ are separated by an energy equivalent to that of $\nu = 9.8$ GHz photons (X-band EPR). The existence of multiple nuclear spin states does not invalidate the proposed operation scheme because of the $\Delta m_{I} = 0$ selection rule.

Concluding, we have shown that molecular clusters containing two lanthanide ions meet the ingredients required to implement a CNOT quantum gate. The definition of control and target qubits is based on the magnetic inequivalence of the two ions, which has been achieved by chemically engineering dissimilar coordination spheres. Although we have restricted here to discussing the [Tb$_{2}$] compound, for which the magnetic asymmetry can be more easily determined on account of its large angular momentum, the same molecular structure can be realized with other members of the lanthanide series.\cite{Aguila10} This flexibility, characteristic of molecular materials, enables a vast choice of qugate designs. For instance, it will be possible to select isotopes with zero nuclear spin (like most of Er isotopes) or ions with dominantly easy-plane anisotropy that, like Ce ($J=5/2$), have a $m_{J} = \pm 1/2$ electronic ground state doublet. In the latter case, the CNOT operation would correspond to an allowed electromagnetic transition, which should enhance the attainable Rabi frequencies to values of order $120$ MHz/mT. Chemically engineered molecular qugates can therefore open promising avenues for the realization of scalable quantum computing architectures.

\begin{acknowledgments}

The present work was partly funded by grants MAT$2009-13977-$C$03$ (MOLCHIP), CTQ$2009-06959$, FIS$2008-01240$, and FIS$2009-13364-$C$02$, from the Spanish MICINN, and the Consolider-Ingenio project on Molecular Nanoscience. Funding from the European Research Council Starting Grant FuncMolQIP (to GA) is also acknowledged. G. A. acknowledges Generalitat de Catalunya for the prize ICREA Academia $2008$.

\end{acknowledgments}

%%%%%%%%%%%%%%%%%%%%%%%%%%%%%%%%%%%%%%%%%%%%%%%%%%%%%%%%%%%%
%%%%%%%%%%%%%%%%%%%%%%   BIBLIOGRAPHY   %%%%%%%%%%%%%%%%%%%%
%%%%%%%%%%%%%%%%%%%%%%   BIBLIOGRAPHY   %%%%%%%%%%%%%%%%%%%%
%%%%%%%%%%%%%%%%%%%%%%   BIBLIOGRAPHY   %%%%%%%%%%%%%%%%%%%%
%%%%%%%%%%%%%%%%%%%%%%   BIBLIOGRAPHY   %%%%%%%%%%%%%%%%%%%%
%%%%%%%%%%%%%%%%%%%%%%   BIBLIOGRAPHY   %%%%%%%%%%%%%%%%%%%%
%%%%%%%%%%%%%%%%%%%%%%%%%%%%%%%%%%%%%%%%%%%%%%%%%%%%%%%%%%%%

%%%%%%%%%%%%%%%%%%%%%%%%%%%%%%%%%%%%%%%%%%%%%%%%%%%%%%%%%%%%
%%%%%%%%%%%%%%%%%%%%%%%%%%%%%%%%%%%%%%%%%%%%%%%%%%%%%%%%%%%%
%%%%%%%%%%%%%%%%%%%%%%%%%%%%%%%%%%%%%%%%%%%%%%%%%%%%%%%%%%%%
%%%%%%%%%%%%%%%%%%%%%%%%%%%%%%%%%%%%%%%%%%%%%%%%%%%%%%%%%%%%
%%%%%%%%%%%%%%%%%%%%%%%%%%%%%%%%%%%%%%%%%%%%%%%%%%%%%%%%%%%%
%%%%%%%%%%%%%%%%%%%%%%%%%%%%%%%%%%%%%%%%%%%%%%%%%%%%%%%%%%%%
%%%%%%%%%%%%%%%%%%%%%%%%%%%%%%%%%%%%%%%%%%%%%%%%%%%%%%%%%%%%


\begin{thebibliography}{10}

\bibitem{Nielsen00} M. A. Nielsen and I. L. Chuang, {\em Quantum computation and quantum information}, Cambridge University Press (New York, 2000).

\bibitem{Ladd10} T. D. Ladd {\em et al.}, Nature {\bf 464}, $45$ (2010).

\bibitem{Clarke08} J. Clarke and F. K. Wilhelm, Nature {\bf 453}, $1031$ (2008).

\bibitem{Plantenberg07} J. H. Plantenberg, P. C. de Groot, C. J. P. M. Harmans, and J. E. Mooij, Nature {\bf 447}, $836$ (2007).

\bibitem{DiCarlo09} L. DiCarlo {\em et al.}, Nature {\bf 60}, 240 (2009).

\bibitem{Burkard99} G. Burkard, D. Loss, and D. P. DiVincenzo, Phys. Rev. B {\bf 59}, $2070$ (1999).

\bibitem{Jelezko04} F. Jelezko, T. Gaebel, I. Popa, A. Gruber,and J. Wrachtrup, Phys. Rev. Lett. {\bf 92}, $76401$ (2004).

\bibitem{Hanson08} R. Hanson and D. D. Awschalom, Nature {\bf 453}, $1043$ (2008).

\bibitem{Leuenberger01}  M. Leuenberger and D. Loss, Nature {\bf 410}, $789$ (2001).

\bibitem{Tejada01} J. Tejada, E. Chudnovsky, E. del Barco, J. Hern\'andez, and T. Spiller, Nanotechnology {\bf 12}, 181 (2001).

\bibitem{Troiani05} F. Troiani, {\em et al.}, Phys. Rev. Lett. {\bf 94}, 207208 (2005).

\bibitem{Lehmann07} J. Lehmann, A. Gaita-Ari\~{n}o, E. Coronado, and D. Loss, Nature Nanotechnology {\bf 2}, $312$ (2007).

\bibitem{Stamp09} P. C. E. Stamp and A. Gaita-Ari\~{n}o, J. Mater. Chem. {\bf 19}, $1718$ (2009).

\bibitem{Ardavan09} A. Ardavan and S. J. Blundell, J. Mater. Chem. {\bf 19},  $1754$ (2009).

\bibitem{Miyake09} K. Miyake {\em et al.}, J. Am. Chem. Soc. {\bf 131}, $17808$ (2009).

\bibitem{Mannini10} M. Mannini {\em et al.}, Nature {\bf 468}, 417 (2010).

\bibitem{Ardavan07} A. Ardavan {\em et al.}, Phys. Rev. Lett. {\bf 98}, 057201 (2007).

\bibitem{Bertaina08} S. Bertaina {\em et al.}, Nature {\bf 453}, 203 (2008).

\bibitem{Schlegel08} C. Schlegel, J. van Slageren, M. Manoli, E. K. Brechin, and M. Dressel, Phys. Rev. Lett. {\bf 101}, 147203(4pp) (2008).

\bibitem{Timco09} G. A. Timco {\em et al.}, Nature Nanotechnology {\bf 4}, 173 (2009).

 \bibitem{Candini09} A. Candini {\em et al.}, Phys. Rev. Lett. {\bf 104}, $037203$ (2010).

\bibitem{Bertaina07}  S. Bertaina {\em et al.}, Nature Nanotechnology {\bf 2}, $39$ (2007).

\bibitem{Aguila10} D. Aguil\`{a} {\em et al.}, Inorg. Chem. {\bf 49}, $6784$ (2010).

\bibitem{EPAPS} See EPAPS Document No. [number will be inserted by publisher] for further details on the structural and physical characterization of the samples, as well as for a theoretical derivation of Eq. (\ref{Hamiltonian1}). For more information on EPAPS, see http://www.aip.org/pubservs/epaps.html.

\bibitem{Martinez10} M. J. Mart\'{\i}nez-P\'erez, J. Ses\'e, F. Luis, D. Drung, and T. Schurig, Rev. Sci. Instrum. {\bf 81}, 016108 (2010).

\bibitem{GarciaPalacios09} J. L. Garcia-Palacios, J. B. Gong , and F. Luis, J. Phys. Condens. Matter {\bf 21}, 456006 (2009).






\end{thebibliography}
\end{document}